\documentclass[twocolumn,aps,showpacs,showkeys,nofootinbib]{revtex4}
\usepackage{epsfig}

\newcommand{\be}{\begin{equation}}
\newcommand{\ee}{\end{equation}}
\newcommand{\ba}{\begin{eqnarray}}
\newcommand{\ea}{\end{eqnarray}}

\def\la{\mathrel{\mathpalette\fun <}}
\def\ga{\mathrel{\mathpalette\fun >}}
\def\fun#1#2{\lower3.6pt\vbox{\baselineskip0pt\lineskip.9pt
        \ialign{$\mathsurround=0pt#1\hfill##\hfil$\crcr#2\crcr\sim\crcr}}}
        
\def\rmFoM{{\rm FoM}}

\def\bfk{\mbox{\bf k}}
\def\bfr{\mbox{\bf r}}
\def\etal{{\frenchspacing\it et al.}}
\def\rmd{\mbox{d}}

\def\rmdet{{\rm det}}
\def\rmCov{{\rm Cov}}
\def\rmFoM{{\rm FoM}}

\begin{document}
\input{epsf.sty}

\title{Clarifying Forecasts of Dark Energy Constraints
from Baryon Acoustic Oscillations }
\author{Yun~Wang}
\address{Homer L. Dodge Department of Physics \& Astronomy, Univ. of Oklahoma,
                 440 W Brooks St., Norman, OK 73019;
                 email: wang@nhn.ou.edu}

                 \today

\begin{abstract}

The measurement of baryon acoustic oscillations (BAO) from a galaxy
redshift survey provides one of the most promising methods for
probing dark energy. In this paper, we clarify the assumptions
that go into the forecasts of dark energy constraints
from BAO. We show that assuming a constant $nP_{0.2}/G^2(z)$ 
(where $P_{0.2}$ is the real space galaxy power spectrum at
$k=0.2\,h/$Mpc and redshift $z$) gives a good approximation of the 
observed galaxy number density expected from a realistic flux-limited 
galaxy redshift survey. We find that assuming $nP_{0.2}/G^2(z)=10$ gives
very similar dark energy constraints to assuming $nP_{0.2}=3$, but
the latter corresponds to a galaxy number density larger 
by $\sim$ 70\% at $z=2$.
We show how the Figure-of-Merit (FoM) for constraining dark energy 
depends on the assumed galaxy number density, redshift accuracy, 
redshift range, survey area, and the systematic errors due to calibration 
and uncertainties in the theory of nonlinear evolution and galaxy 
biasing. We find that an additive systematic noise of up to 
$0.4-0.5$\% per $\Delta z=0.1$ redshift slice does not lead to 
significant decrease in the BAO FoM.

\end{abstract}

\pacs{98.80.Es,98.80.-k,98.80.Jk}

\keywords{Cosmology}

\maketitle

\section{Introduction}

It has been more than a decade since the discovery of the cosmic 
acceleration \cite{Riess98,Perl99}. Illuminating the nature 
of dark energy, the unknown cause of the observed cosmic 
acceleration, has become a prominent task for the astronomy
and physics communities.

Dark energy could be an unknown energy component \cite{quintessence},
or a modification of general relativity \cite{modifiedgrav}.
These two classes of models can be differentiated 
if both the cosmic expansion history and growth history of
cosmic large scale structure are accurately and precisely
measured \cite{DEvsMG}.
Current observational data are consistent with dark energy
being a cosmological constant, but other explanations
are still allowed (see, for example, \cite{data}).
See \cite{reviews} for recent reviews on dark energy research.

One of the most promising methods for probing dark energy is
to use the baryon acoustic oscillations (BAO) in the observed
3-D galaxy distribution as a cosmological standard ruler
\cite{BG03,SE03,BAO}. At the last scattering of cosmic
microwave background (CMB) photons, the acoustic oscillations 
in the photon-baryon fluid became frozen, and imprinted their signatures
on both the CMB (the acoustic peaks in the CMB angular
power spectrum) and the matter distribution (the BAO
in the galaxy power spectrum).
Because baryons comprise only a small fraction of matter,
and the matter power spectrum has evolved significantly
since last scattering of photons, BAO are much smaller in amplitude
than the CMB acoustic peaks, and are washed out on small
scales. BAO in the observed galaxy power spectrum have the characteristic 
scale determined by the comoving sound horizon at recombination, 
which is precisely measured by the CMB anisotropy data \cite{Spergel07}.
In principle, the BAO scale can be extracted from data
both in the transverse direction ($s_{\perp}$), and
along the line-of-sight ($s_{\parallel}$).
Comparing the observed BAO scales with the expected values gives
the angular diameter distance $D_A(z)=r(z)/(1+z)$ (where
$r(z)$ is the comoving distance) in the transverse 
direction, and the Hubble parameter $H(z)$ in the radial direction:
\ba
s_{\perp} &\propto & \frac{s}{D_A(z)} \nonumber\\
s_{\parallel} &\propto & s H(z).
\ea

Seo \& Eisenstein (2007) \cite{SE07} (henceforth SE07) provided simple 
fitting formulae for forecasting the accuracies of $s/D_A(z)$ and $s\,H(z)$
from future galaxy redshift surveys. In this paper, we will clarify the 
assumptions made in BAO forecasts, and investigate their implications.
We will discuss the method in Sec.2, present 
results on BAO forecasts in Sec.3, and summarize in Sec.4.

\section{The method}

The simplest and most widely used method to forecast constraints from 
future observations is to use the Fisher matrix formalism.
The Fisher information matrix of a given set of parameters, ${\bf s}$,
approximately quantifies the amount of
information on ${\bf s}$ that we expect to get from our future data.
The Fisher matrix can be written as
\be
\label{eq:fisherdef}
F_{ij}=- \frac{\partial ^2 \ln\,L}{\partial s_i \partial s_j},
\ee
where $L$ is the likelihood function, 
the expected probability distribution of the observables given
parameters ${\bf s}$. The Cram{\'e}r-Rao inequality states that no 
unbiased method can measure the $i$-th parameter with standard
deviation less than $1/\sqrt{F_{ii}}$ if other parameters are known,
and less than $\sqrt{ ({\bf F}^{-1})_{ii} }\,$ if other parameters
are estimated from the data as well \cite{Kendall69}.

In the limit where the length scale corresponding to
the survey volume is much larger than
the scale of any features in the galaxy power spectrum $P_g(k)$, 
we can assume that the likelihood function for the band powers of a 
galaxy redshift survey is Gaussian \cite{Feldman94}. 
Then the Fisher matrix for estimating parameters from a
galaxy redshift survey can be approximated as \citep{Tegmark97}
\be
F_{ij}= \int_{k_{min}}^{k_{max}}
\frac{\partial\ln P_g(\bfk)}{\partial p_i}
\frac{\partial\ln P_g(\bfk)}{\partial p_j}\,
V_{eff}(\bfk)\, \frac{d \bfk^3}{2\, (2\pi)^3}
\label{eq:full Fisher}
\ee
where $p_i$ are the parameters to be estimated from data, and 
the derivatives are evaluated at parameter values of the
fiducial model. The effective volume of the survey
\ba
V_{eff}(k,\mu) &=&\int \rmd^3\bfr \left[ \frac{n(\bfr) P_g(k,\mu)}
{ n(\bfr) P_g(k,\mu)+1}\right]^2 \nonumber\\
&=&\left[ \frac{ n P_g(k,\mu)}{n P_g(k,\mu)+1} \right]^2 V_{survey},
\ea
where the comoving number density $n$ is assumed to only depend on
the redshift for simplicity; $\mu = \bfk \cdot \hat{\bfr}/k$, 
with $\hat{\bfr}$ denoting the unit vector along the line of sight; 
$\bfk$ is the wavevector with $|\bfk|=k$. 
Note that the Fisher matrix $F_{ij}$ is 
the inverse of the covariance matrix of the parameters $p_i$ if the 
$p_i$ are Gaussian distributed. Eq.(\ref{eq:full Fisher}) propagates the 
measurement errors in $\ln P_g(\bfk)$ (which are proportional to 
$[V_{eff}(\bfk)]^{-1/2}$) into measurement errors for the parameters $p_i$.
Note that Eq.(\ref{eq:full Fisher}) can be rewritten as
\ba
F_{ij}&=& V_{survey} \int_{-1}^{1}d\mu
\int_{k_{min}}^{k_{max}}
\frac{\partial P_g(k,\mu)}{\partial p_i}
\frac{\partial P_g(k,\mu)}{\partial p_j}\,\cdot\nonumber\\
& & \cdot \left[ \frac{1}{P_g(k,\mu)+n^{-1}}\right]^2
 \frac{2\pi k^2 dk}{2\, (2\pi)^3},
\label{eq:Fisher2}
\ea
where $\mu=\hat{\bfk} \cdot \hat{\bfr}$.

\subsection{The ``wiggles only'' fitting formulae}

SE07 \cite{SE07} provided simple fitting formulae
for forecasting the accuracies of $s/D_A(z)$ and $s\,H(z)$
from future galaxy redshift surveys.
Essentially, they approximated Eq.(\ref{eq:Fisher2})
with
\ba
F_{ij}&\simeq& V_{survey} \int_{-1}^{1}d\mu
\int_{k_{min}}^{k_{max}}
\frac{\partial P_b(k,\mu|z)}{\partial p_i}
\frac{\partial P_b(k,\mu|z)}{\partial p_j}\,\cdot\nonumber\\
& & \cdot \left[ \frac{1}{P_g^{lin}(k,\mu|z)+n^{-1}}\right]^2
 \frac{2\pi k^2 dk}{2\, (2\pi)^3},
\label{eq:Fisher3_rev}
\ea
where $P_b(k,\mu|z)$ is the power spectrum that contains baryonic features.
The linear galaxy power spectrum 
\ba
\label{eq:Pk}
P_g^{lin}(k,\mu|z)&=& P_{g,r}^{lin}(k|z)\,R(\mu)\\
P_{g,r}^{lin}(k|z)&=&b(z)^2 \left[\frac{G(z)}{G(0)}\right]^2\, P_m^{lin}(k|z=0)
\ea
where $P_{g,r}^{lin}(k|z)$ is the linear galaxy power spectrum in real space,
$b$ is the bias factor, $G(z)$ is the growth factor, and
$P_m^{lin}(k|z=0)$ is the present day linear matter power spectrum.
$R(\mu)$ is the linear redshift distortion factor given by 
\be
R(\mu)=\left(1 + \beta \mu^2 \right)^2.
\ee

The power spectrum that contains baryonic features, 
$P_b(k,\mu)$, is given by
\ba
\label{eq:Pb_rev}
P_b(k,\mu|z)&=& \sqrt{8\pi^2} A_0 \,P_g^{lin}(k_{0.2},\mu|z) 
\frac{\sin (x)}{x} \cdot\nonumber\\
&& \cdot \exp\left[ -(k \Sigma_s)^{1.4} -\frac{k^2\Sigma_{nl}^2}{2}\right],
\ea
where we have define
\ba
k_{0.2} &\equiv&  0.2\,h\,{\rm Mpc}^{-1}\\
x &\equiv&  \left(k^2_\perp s ^2_\perp +k^2_\parallel s^2_\parallel\right)^{1/2}
\\
k_\parallel &=& \bfk \cdot \hat{\bfr}=k\mu \\
k_\perp &= &\sqrt{k^2-k^2_\parallel}=k\sqrt{1-\mu^2}.
\ea
The nonlinear damping scale 
\ba
\label{eq:NL}
\Sigma_{nl}^2&=&(1-\mu^2) \Sigma_{\perp}^2+\mu^2 \Sigma^2_{\parallel}
\nonumber\\
\Sigma_{\parallel}&=&\Sigma_{\perp} (1+f_g) \nonumber\\
\Sigma_{\perp}&= &12.4 \,h^{-1}{\rm Mpc}\, \left(\frac{\sigma_8}{0.9}\right)
\cdot 0.758\, \frac{G(z)}{G(0)}\, p_{NL}\nonumber\\
&=& 8.355 \,h^{-1}{\rm Mpc}\, \left(\frac{\sigma_8}{0.8}\right)
\cdot \frac{G(z)}{G(0)} \, p_{NL},
\ea
where $f_g=d\ln G(z)/d\ln a$ denotes the growth rate of 
matter density fluctuations. The parameter $p_{NL}$ 
indicates the remaining level
of nonlinearity in the data; with $p_{NL}=0.5$ (50\% nonlinearity)
as the best case, and $p_{NL}=1$ (100\% nonlinearity)
as the worst case \cite{SE07}. For a fiducial
model based on WMAP3 results \cite{Spergel07}
($\Omega_m=0.24$, $h=0.73$, $\Omega_{\Lambda}=0.76$, $\Omega_k=0$,
$\Omega_b h^2=0.0223$, $\tau=0.09$, $n_s=0.95$, $T/S=0$),
$A_0=0.5817$, $P_{0.2}=2710\, \sigma_{8,g}^2$, the
Silk damping scale $\Sigma_s=8.38\,h^{-1}{\rm Mpc}$.

Defining 
\ba
p_1 &=& \ln s_\perp^{-1} =\ln (D_A/s), \\
p_2 &=& \ln s_\parallel=\ln(s H),
\ea
substituting Eq.(\ref{eq:Pb_rev}) into Eq.(\ref{eq:Fisher3_rev}),
and making the approximation of $\cos^2 x \sim 1/2$, we find
\ba
&&\hspace{-0.1 in} F_{ij}\simeq V_{survey} A_0^2 \int_0^{1}\rmd\mu \,f_i(\mu) \,f_j(\mu)
\int_0^{k_{max}} \rmd k \,k^2 \cdot \nonumber \\
& & \hskip 0.02in \cdot \left[\frac{ P^{lin}_m(k|z=0)}
{ P^{lin}_m(k_{0.2}|z=0)} +
\frac{1}{n P^{lin}_g(k_{0.2},\mu|z) \,e^{-k^2\mu^2\sigma_r^2}}\right]^{-2} 
\nonumber \\
& & \hskip 0.02in \cdot \exp\left[ -2(k \Sigma_s)^{1.4} -k^2\Sigma_{nl}^2\right],
\label{eq:Fisher_Wang}
\ea
where $P_g^{lin}(k_{0.2},\mu|z)$ is given by Eq.(\ref{eq:Pk}) with $k=k_{0.2}$.
Note that we have added the damping factor, $e^{-k^2\mu^2\sigma_r^2}$, 
due to redshift uncertainties, with
\be
\sigma_r = \frac{\partial r}{\partial z}\, \sigma_z
\ee
where $r$ is the comoving distance. For a flat universe,
$r=\int_0^z dz'/H(z')$.
The functions $f_i(\mu)$ are given by
\ba
f_1(\mu)&=&\partial \ln x/\partial \ln p_1=\mu^2-1\\
f_2(\mu)&=&\partial \ln x/\partial \ln p_2=\mu^2.
\ea

\subsection{Full power spectrum calculation}

In order to compare the ``wiggles only'' method and
the full power spectrum method for BAO forecast, we must
include the nonlinear effects in the same way in both methods. 
In the full power spectrum method, 
the observed power spectrum is reconstructed using a particular reference 
cosmology, including the effects of bias and redshift-space distortions \cite{SE03}:
\ba
\label{eq:P(k)}
P_{obs}(k^{ref}_{\perp},k^{ref}_{\parallel}) &=&
\frac{\left[D_A(z)^{ref}\right]^2  H(z)}{\left[D_A(z)\right]^2 H(z)^{ref}}
\, b^2 \left( 1+\beta\, \mu^2 \right)^2
\cdot \nonumber\\
& & \hskip -0.3in \cdot\left[ \frac{G(z)}{G(0)}\right]^2 P_{matter}(k|z=0)+ P_{shot},
\ea
where 
\be
k^{ref}_{\perp}=k_\perp\,\frac{ D_A(z)}{D_A(z)^{ref}}, \hskip 0.5cm
k^{ref}_{\parallel}=k_\parallel\,\frac{H(z)^{ref}}{H(z)},
\ee
and $\mu^2=k^2_{\parallel}/k^2=k^2_{\parallel}/(k^2_{\perp}+k^2_{\parallel})$.
The values in the reference cosmology are denoted by the subscript ``ref'',
while those in the true cosmology have no subscript.

Following Seo \& Eisenstein (2007) \cite{SE07}, we include
nonlinear effects in the full power spectrum calculation
by modifying the derivatives of $P(\bfk)$ with respect to
the parameters $p_i$, i.e.,
\be
\frac{\partial P_g(k,\mu|z)}{\partial p_i}=
\frac{\partial P_g^{lin}(k,\mu|z)}{\partial p_i}\cdot
\exp\left(-k^2\Sigma_{nl}^2/2\right).
\ee
Eq.(\ref{eq:Fisher2}) becomes
\ba
F_{ij}&=& V_{survey} \int_{-1}^{1}d\mu
\int_{k_{min}}^{k_{max}}
\frac{\partial \ln P_g^{lin}(k,\mu)}{\partial p_i}
\frac{\partial \ln P_g^{lin}(k,\mu)}{\partial p_j}\,\cdot\nonumber\\
& & \cdot \left[ \frac{n P_g^{lin}(k,\mu)}{nP_g^{lin}(k,\mu)+1}\right]^2
\, e^{-k^2\Sigma_{nl}^2} \frac{2\pi k^2 dk}{2\, (2\pi)^3}.
\label{eq:Fisher2_nl}
\ea

\subsection{Figure of merit and dark energy parameters}
\label{sec:FoM}

The DETF defined the dark energy figure of merit (FoM) to be
the inverse of the area enclosed by the 95\% confidence contour
for the parameters ($w_0,w_a$), assuming a dark energy equation of state
$w_X(z) =w_0+(1-a) w_a$ \cite{Chev01}. It is most convenient to
define a relative generalized FoM \cite{Wang08}
\be
\rmFoM_r=\frac{1}{\sqrt{ \rmdet\,\rmCov(f_1,f_2,f_3,...)}},
\label{eq:FoM_r}
\ee
where $\{f_i\}$ are the chosen set of dark energy parameters.
This definition has the advantage of being easy to calculate for
either real or simulated data. When applied to ($w_0,w_a$), we find
\be
\rmFoM_r=\frac{1}{\sqrt{ \rmdet\,\rmCov(w_0,w_a)}}
=\frac{1}{\sqrt{\sigma^2_{w_0}\sigma^2_{w_a}-\sigma^2_{w_0,w_a}}},
\ee
which differs by a factor of 6.17$\pi$ from the DETF definition.
This is the FoM that has been widely used, and tabulated in the DETF 
report.

The most sensible FoM requires choice of dark energy parameters
that are least correlated \cite{Wang08}. A good choice is
to use ($w_0,w_{0.5}$) from
\ba
\label{eq:w_0.5}
&&w_X(a)=3 w_{0.5}-2 w_0 + 3\left( w_0 - w_{0.5} \right) \,a\\
&&X(z)= (1+z)^{3(1-2 w_0+ 3w_{0.5})} \exp\left[9(w_0-w_{0.5})\, \frac{z}{1+z}\right],
\nonumber
\ea
where $w_{0.5}$ is the value of $w_X$ at $z=0.5$.
The correlation of $(w_0,w_{0.5})$ is much smaller than that of $(w_0,w_a)$.
For real data analyzed using Marcov Chain Monte Carlo (MCMC), the covariance 
matrix of $(w_0,w_{0.5})$ and $(w_0,w_a)$ cannot be transformed, since 
choosing $(w_0,w_{0.5})$ and choosing $(w_0,w_a)$ as the base parameters
correspond to different priors, if uniform priors are assumed for
the base parameters. For Fisher matrix forecast, one can simply
transform the covariance matrix of $(w_0,w_a)$ into that of
$(w_0,w_{0.5})$ by using
\be
w_{0.5}=w_0 + w_a/3.
\ee

\subsection{Choice of redshift bins}

There has been some confusion in the literature about 
the optimal choice of redshift bins in constraining dark energy.
This choice should depend on the observational method considered.
For a galaxy redshift survey, the observables are
$s/D_A$ and $sH$ (length scales extracted from data analysis).
Since these scales are assumed to be constant in each redshift
slice, the redshift slices should be chosen such that the 
variation of $1/D_A(z)$ and $H(z)$ in each redshift slice 
remain roughly constant with $z$.

For a flat universe, the variations of $H(z)$ and $1/D_A(z)$ with $z$ are
\ba
\frac{d(H/H_0)}{dz}&=&\frac{3\Omega_m(1+z)^2+\Omega_X X'(z)}{2E(z)}\\
\frac{d(1/D_A)}{dz}&=& \frac{1}{r(z)} \left[1-\frac{1+z}{H(z)\,r(z)}\right]\\
E^2(z)&=& \frac{H^2(z)}{H_0^2}=\Omega_m(1+z)^3+\Omega_X X(z),
\ea
where $X(z)\equiv \rho_X(z)/\rho_X(0)$ is the dark energy density function.
$X(z)=1$ for a cosmological constant.
Fig.\ref{fig:dz} shows the variations of $H(z)$ and $1/D_A(z)$ with $z$
for a fiducial flat $\Lambda$CDM model. The amplitude of variation
in $H(z)$ {\it increases} slowly with $z$. The amplitude of variation
in $1/D_A(z)$ {\it decreases} with $z$, and stablize for $z\ga 1$.
{\emph{Thus choosing a constant $\Delta z$ for redshift slices
is the optimal choice.} Choosing constant $\Delta z$ avoids significantly 
degrading the approximation of $H(z)$ being constant in each redshift slice
as $z$ increases, while the approximation of $D_A(z)$ being constant becomes 
increasingly better as $z$ increases. This makes sense since $H(z)$
measurements carry more weight in constraining dark energy than
the $D_A(z)$ measurements.

\begin{figure} 
\psfig{file=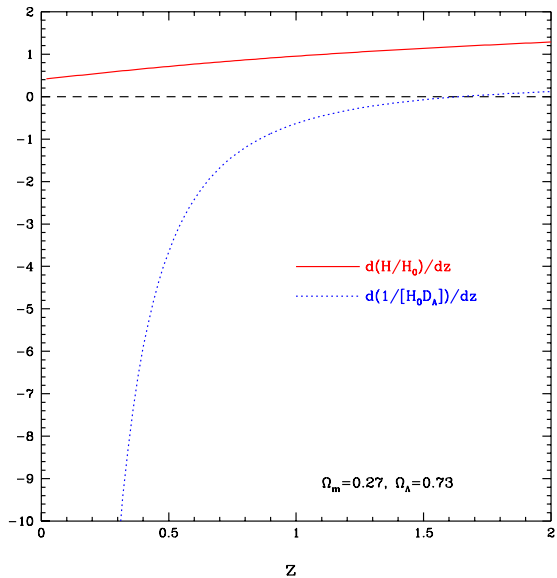,width=3.5in}
\caption{\label{fig:dz}
\footnotesize%
The variations of $H(z)$ and $1/D_A(z)$ with $z$.
}
\end{figure}

\section{Results}

We find that the fitting formulae, Eq.(\ref{eq:Fisher_Wang}),
give errors of $\ln (s/D_A)$ and $\ln (sH)$ that match
those from the full Fisher matrix, Eq.(\ref{eq:Fisher2_nl}), to
better than $\sim $ 10-20\%. 
We have applied the fitting formulae, Eq.(\ref{eq:Fisher_Wang}),
to investigate the dependence of the forecast of dark energy constraints 
from BAO on various survey parameters. We adopt the Dark Energy Task Force 
(DETF) fiducial model, with $\Omega_m h^2=0.146$, $\Omega_b h^2=0.024$, 
$h=0.725$, $\Omega_k=0$ $w=-1$, and $n_s=1.0$ \cite{detf}. 
We assume $\sigma_{8,m}(z=0)=0.8$ \cite{Komatsu08}, and $\sigma_{8,g}(z=0)=1$. 
We compute the matter transfer function using CMBFAST 4.5.1 \cite{CMBFAST}.
Table 1 lists the parameter values in the DETF model and the ``other'' model
for comparison (all are present day values). We take 
$k_{min}=10^{-4}h\,$Mpc$^{-1}$, and $k_{max}=0.5\,h\,$Mpc$^{-1}$ \cite{SE07}.

\begin{table*}[htb]
\caption{Two fiducial cosmological models}
\begin{center}
\begin{tabular}{l|llllllll}
\hline
name & $\Omega_m$ & $\Omega_{\Lambda}$ & $h$ & ${\Omega_b}h^2$ & $\tau$ & $n_s$ &
 $\sigma_{8,m}$ & $\sigma_{8,g}$\\
\hline
DETF &.2778& .7222&  .725&  .024&  .05&  1&  .8&  1\\
other &.24&  .76&  .73&  .0223&  .09&  .95&  .761&  1\\
\hline
\end{tabular}
\end{center}
\end{table*}

\subsection{Dependence on galaxy number density}

For a given BAO survey, the galaxy number density $n(z)$ and bias
function $b(z)$ should be modeled using available data and supplemented
by cosmological N-body simulations that include galaxies \cite{Angulo08}.
Since $n(z)$ and $b(z)$ depend on survey parameters such as
the flux limit and the target selection method, a more
generic galaxy number density given by assuming
$nP_{0.2}\equiv n P_g^r(k_{0.2}|z)=\,$constant is often used in BAO forecasts, 
where $P_g^r(k_{0.2}|z)$ is the real space power spectrum of galaxies at 
$k=0.2\,h\,$Mpc$^{-1}$ and redshift $z$. 
Fig.\ref{fig:nV} shows the number of galaxies per 
$\Delta z=0.1$ slice per unit volume and per (deg)$^2$ respectively,
multiplied by $b^2(z)$. Clearly, assuming a constant $nP_{0.2}$
corresponds to assuming that $n(z) b^2(z) $ increases faster than a linear
function in $z$, most likely faster than the increase in the galaxy bias 
function $b(z)$ \cite{Sumi09,Orsi09}, thus implying an assumed galaxy number 
density $n(z)$ that {\it increases} with redshift.

\begin{figure} 
\psfig{file=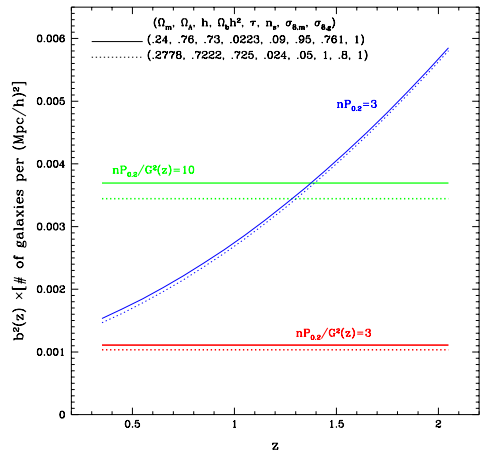,width=2.8in}
\psfig{file=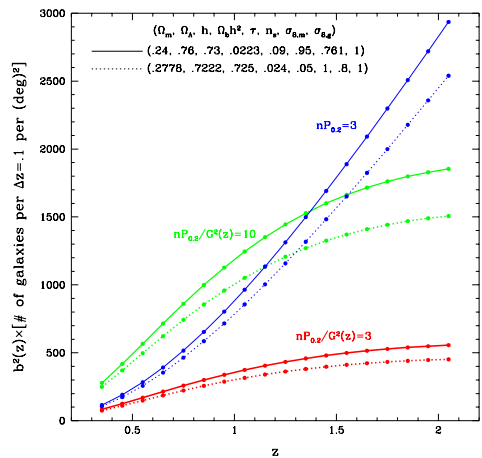,width=2.8in}
\caption{\label{fig:nV}
\footnotesize%
The comoving number density of galaxies per unit comoving volume
(upper panel) and per (deg)$^2$ (lower panel).
The results for two different fiducial cosmological models
are shown, the ``other'' model (solid), and the DETF model (dotted).}
\end{figure}

A typical flux-limited galaxy redshift survey
has a galaxy number density distribution that peaks at some 
intermediate redshift (as the volume per redshift slice
increases with $z$), and decreases toward the high $z$ end of the survey
(as the increase in the number of galaxies fainter than the flux limit
dominates over the increase in volume per redshift slice).
For a galaxy redshift survey with limited resources, photometric 
pre-selection could be used to alter $n(z)$ to cut out low redshift 
galaxies, and validate the assumption of constant $nP_{0.2}$.
However, this would correspond to $nP_{0.2}<1$. 
It is challenging to achieve $nP_{0.2}>1$ at $z=2$ even for the
most optimistic space-based galaxy redshift survey \cite{SPACE,Geach09}.
Thus it is misleading to use $nP_{0.2}=3$ to forecast the
dark energy constraints from a space-based survey.

We propose the following alternative assumption about 
the observed galaxy number density:
\be
nP_{0.2}/G^2(z)=n\, b^2(z) P_m(k_{0.2}|z=0)=\mbox{constant}.
\ee
Fig.\ref{fig:nP} shows the expected $nP_{0.2}$ assuming constant 
$nP_{0.2}/G^2(z)$ in the DETF model. Assuming a constant
$nP_{0.2}/G^2(z)$ allows us to avoid assuming an unrealistically
high $nP_{0.2}$ at the high redshift end, while at the same time
retain the increasingly higher $nP_{0.2}$ easily achievable at
intermediate and lower redshifts that play a key role
in tightening the dark energy constraints.

\begin{figure} 
\psfig{file=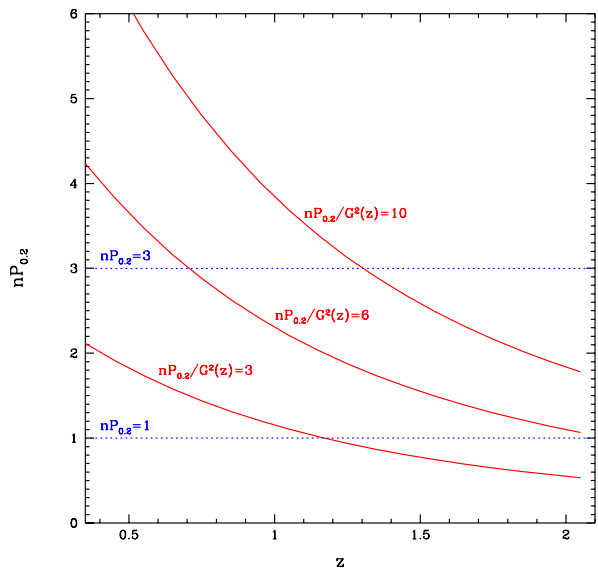,width=3.5in}
\caption{\label{fig:nP}
\footnotesize%
The expected $nP_{0.2}$ assuming constant $nP_{0.2}/G^2(z)$
in the DETF model.}
\end{figure}

In forecasting dark energy FoM from BAO, it is important to specify whether
a constant $nP_{0.2}$ or constant $nP_{0.2}/G^2(z)$ is assumed.
Fig.\ref{fig:DAHz2} shows the fractional errors of $(sH)$ and $(s/D_A)$ 
per $\Delta z=0.1$ slice expected from a 28,000
(deg)$^2$ galaxy redshift survey with $\sigma_z/(1+z)=0.001$, using
the ``wiggles only'' fitting formulae of Eq.(\ref{eq:Fisher_Wang}),
for different assumptions about the observed galaxy number density.
Note that for the ``other'' model, our results are in excellent agreement 
with the results of SE07 \cite{SE07} for $nP_{0.2}=3$
(thin lines with squares, compare with the dashed lines in Fig.3 of SE07),
and for the cosmic variance case, i.e., zero shot noise and zero nonlinearity  
(thin lines with triangles, compare with the solid lines in Fig.3 of SE07).

\begin{figure} 
\psfig{file=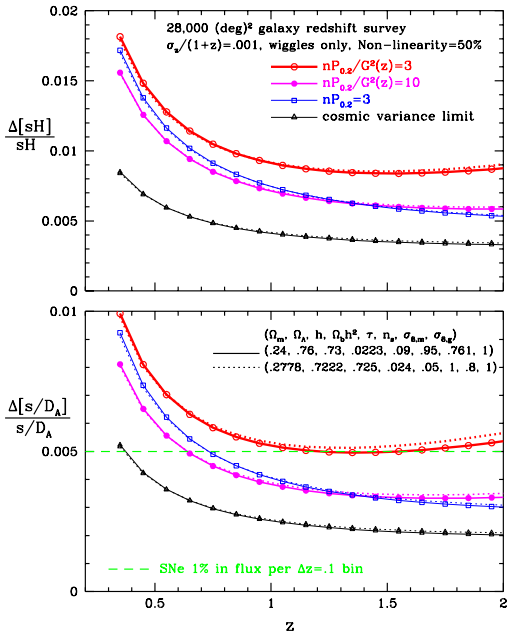,width=3.5in}
\caption{\label{fig:DAHz2}
\footnotesize%
The fractional errors of $(sH)$ and $(s/D_A)$ per $\Delta z=0.1$ slice
expected from a 28,000 (deg)$^2$ galaxy redshift survey with 
$\sigma_z/(1+z)=0.001$, using our ``wiggles only'' modified formula 
of Eq.(\ref{eq:Fisher_Wang}), for different assumptions about the observed 
galaxy number density
The results for two different fiducial cosmological models
are shown, the ``other'' model (solid), and the DETF model (dotted).}
\end{figure}

Fig.\ref{fig:w0wa} shows the 68.3\% joint confidence contours for ($w_0, w_a$)
and ($w_0, w_{0.5}$) (see Sec.\ref{sec:FoM} for the definition of the parameters) 
for the three different galaxy number densities shown in Fig.\ref{fig:nV} and 
Fig.\ref{fig:DAHz2}. The DETF fiducial model is assumed. 
It is interesting to note that assuming $nP_{0.2}/G^2(z)=10$ gives
very similar dark energy constraints to assuming $nP_{0.2}=3$, but
the latter corresponds to a galaxy number density larger by $\sim$ 70\% at 
$z=2$ (see Fig.\ref{fig:nP}).

\begin{figure} 
\psfig{file=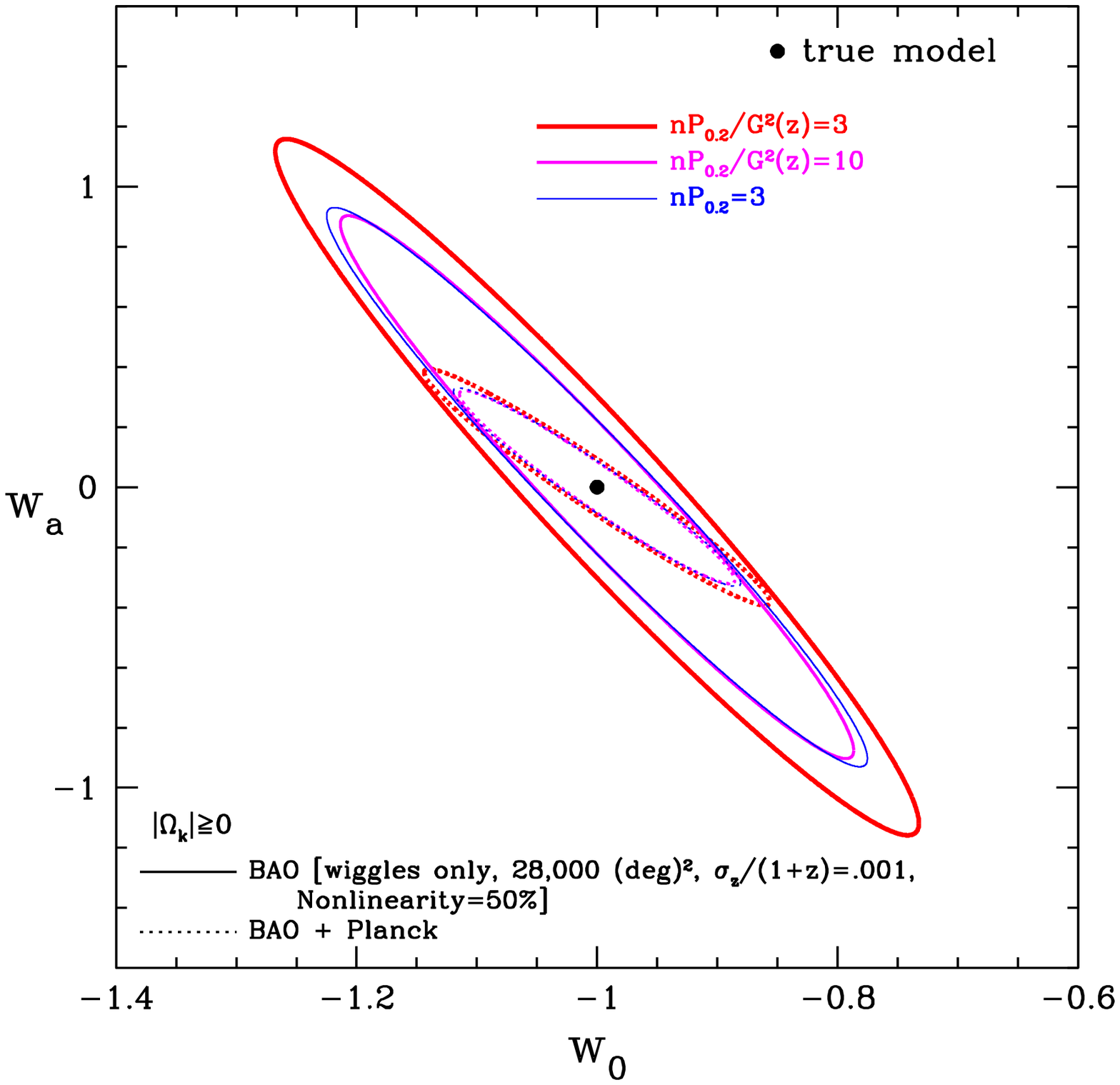,width=2.8in}
\psfig{file=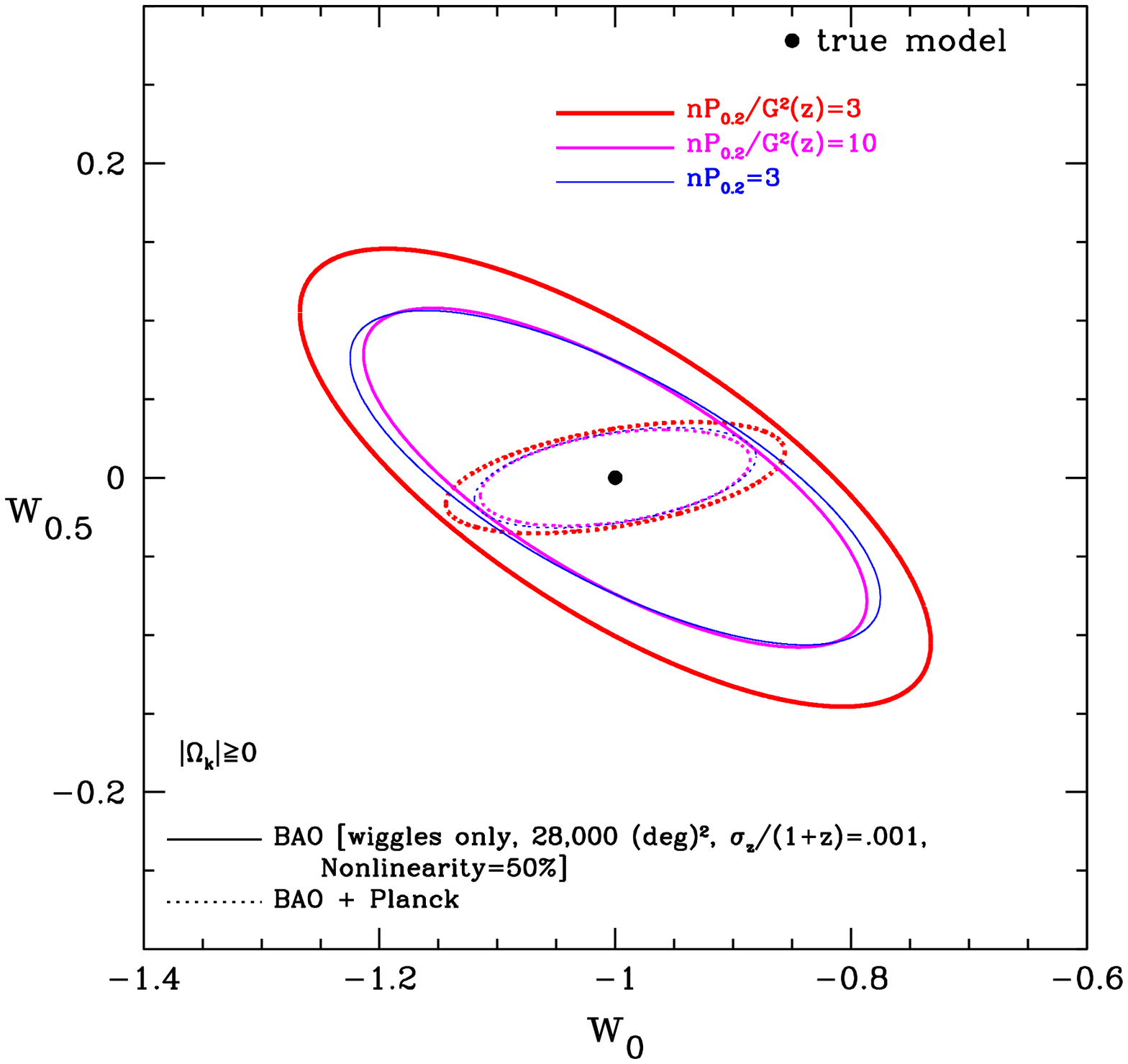,width=2.8in}
\caption{\label{fig:w0wa}
\footnotesize%
The 68.3\% joint confidence contours for ($w_0, w_a$)
and ($w_0, w_{0.5}$) for the three different galaxy number densities
shown in Fig.\ref{fig:nV} and Fig.\ref{fig:DAHz2}.
The DETF fiducial model is assumed.
}
\end{figure}

Fig.\ref{fig:FoM_nP0} shows the relative dark energy FoM, FoM$_r$ (defined
in Sec.\ref{sec:FoM}), for a galaxy redshift survey covering 28,000 (deg)$^2$ 
and $0.3<z<2.1$, as a function of galaxy number density.
The dashed lines in Fig.\ref{fig:FoM_nP0} show the results
of using the full $P(k)$ method (see Eq.[\ref{eq:Fisher2_nl}]).
The full $P(k)$ method boosts the FoM by a factor of $\sim$3-4,
with more gain for higher galaxy number densities.

\begin{figure} 
\psfig{file=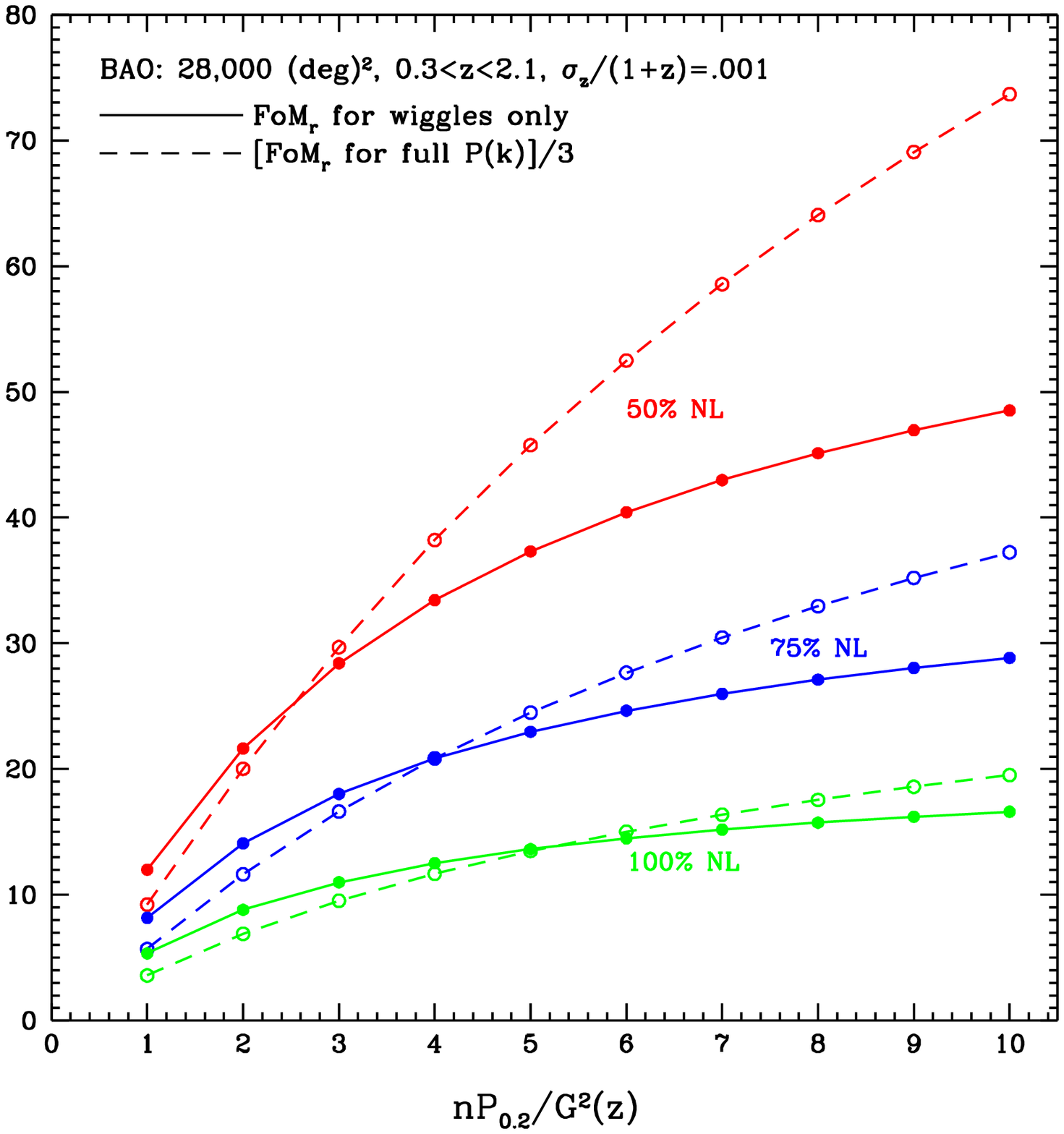,width=3.5in}
\caption{\label{fig:FoM_nP0}
\footnotesize%
The relative dark energy FoM, FoM$_r$, for a galaxy redshift survey
covering 28,000 (deg)$^2$ and $0.3<z<2.1$, as a function of galaxy
number density. Both BAO wiggles only method (solid lines) and
the full $P(k)$ method (dashed lines) are shown.
The DETF fiducial model is assumed.
}
\end{figure}

\subsection{Dependence on other survey parameters}

We now study the dependence of the dark energy FoM on
the other survey parameters, redshift accuracy, redshift range,
and survey area. We show all our results for two representative
galaxy number densities, $nP_{0.2}/G^2(z)=3$ and 10.

Fig.\ref{fig:FoM_dlnz} shows the FoM$_r$ for a galaxy redshift survey
covering 28,000 (deg)$^2$ and $0.3<z<2.1$, as a function of the 
redshift accuracy. The FoM decreases quickly as $\sigma_z/(1+z)$ 
increases beyond 0.001. Increasing the redshift accuracy to
$\sigma_z/(1+z)<0.001$ does not have a significant impact.

\begin{figure} 
\psfig{file=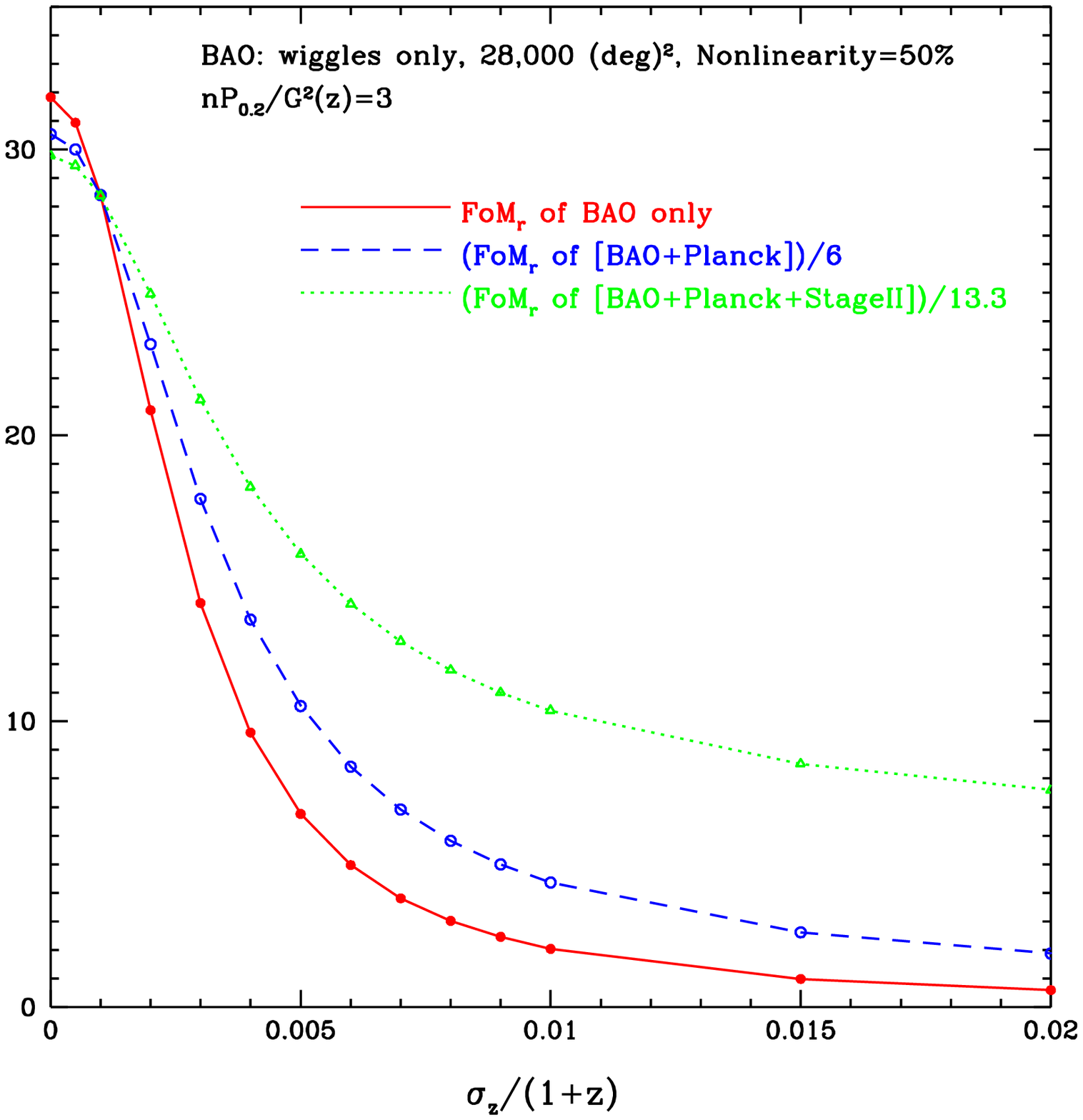,width=2.5in}
\psfig{file=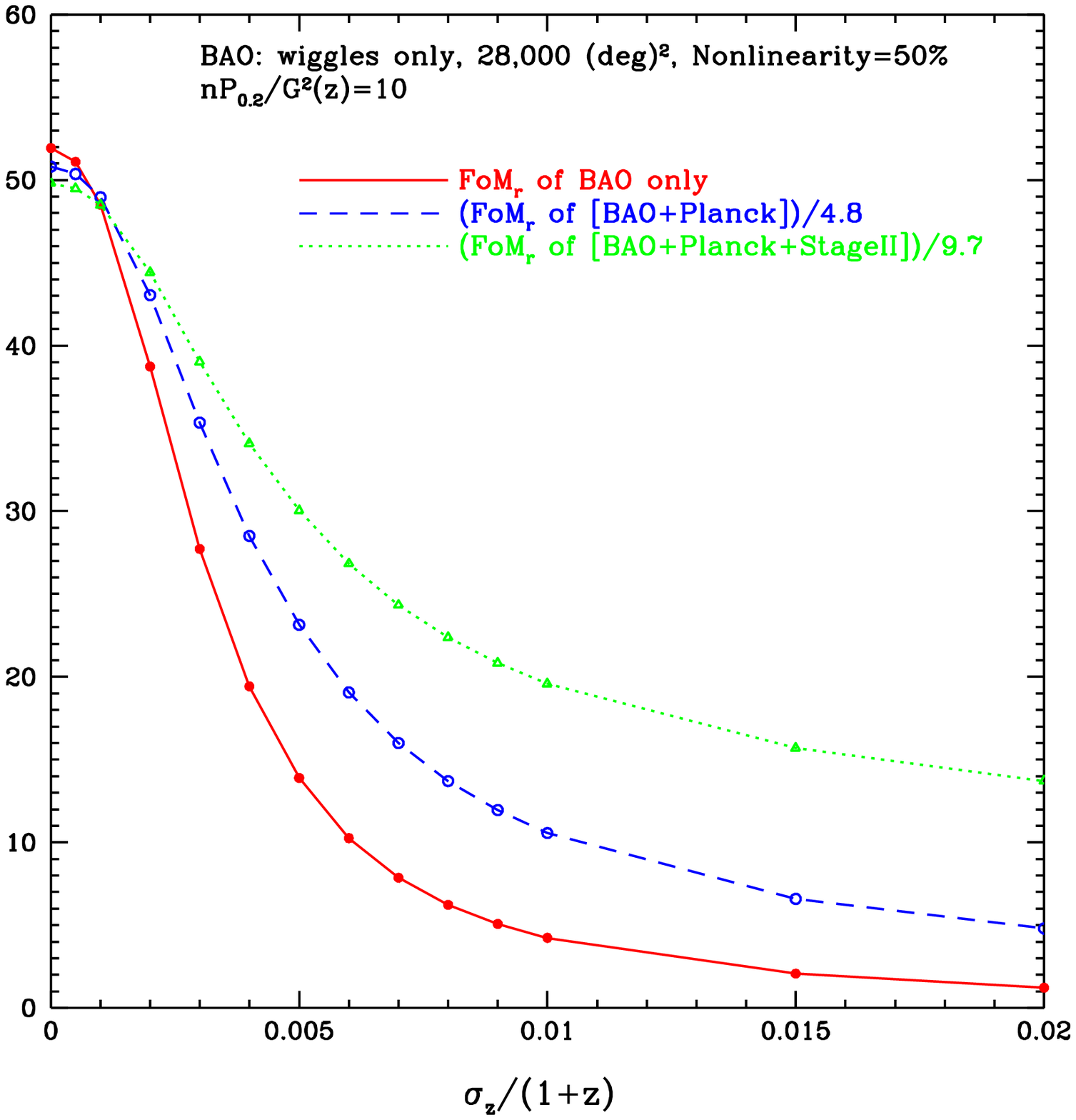,width=2.5in}
\caption{\label{fig:FoM_dlnz}
\footnotesize%
The relative dark energy FoM, FoM$_r$, for BAO wiggles only from
a galaxy redshift survey covering 28,000 (deg)$^2$ and $0.3<z<2.1$, 
as a function of the redshift accuracy.
The DETF fiducial model is assumed.
}
\end{figure}

Fig.\ref{fig:FoM_z} shows FoM$_r$ for a galaxy redshift survey
covering 28,000 (deg)$^2$ and $\sigma_z/(1+z)=.001$, as a function of 
$z_{max}$ for $z_{min}=0.5$ (upper panel), and as a function
of $z_{min}$ for $z_{max}=2$ (lower panel). Decreasing $z_{max}$
and increasing $z_{min}$ both significantly decrease the FoM
for BAO only. 

\begin{figure} 
\psfig{file=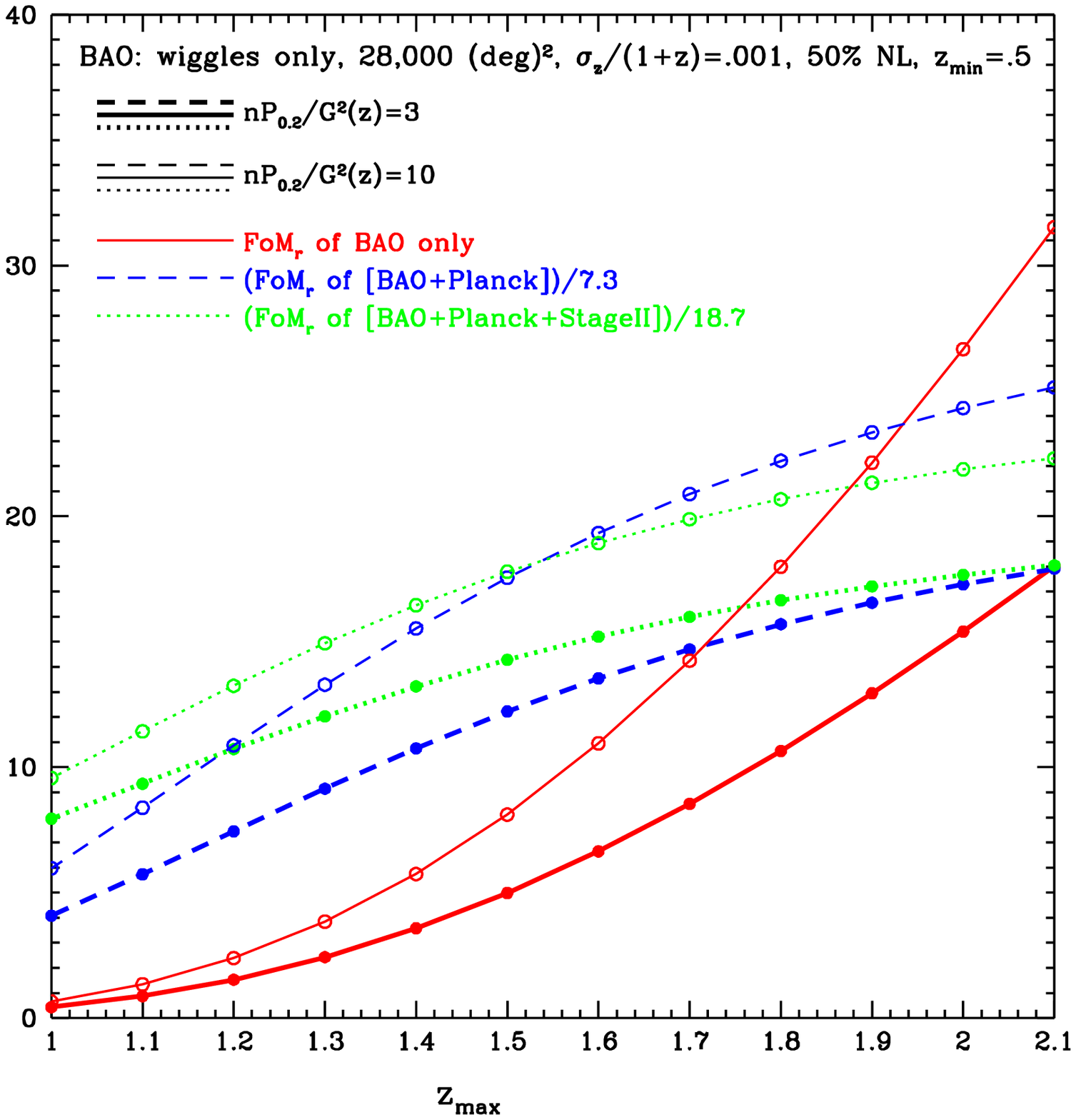,width=2.5in}
\psfig{file=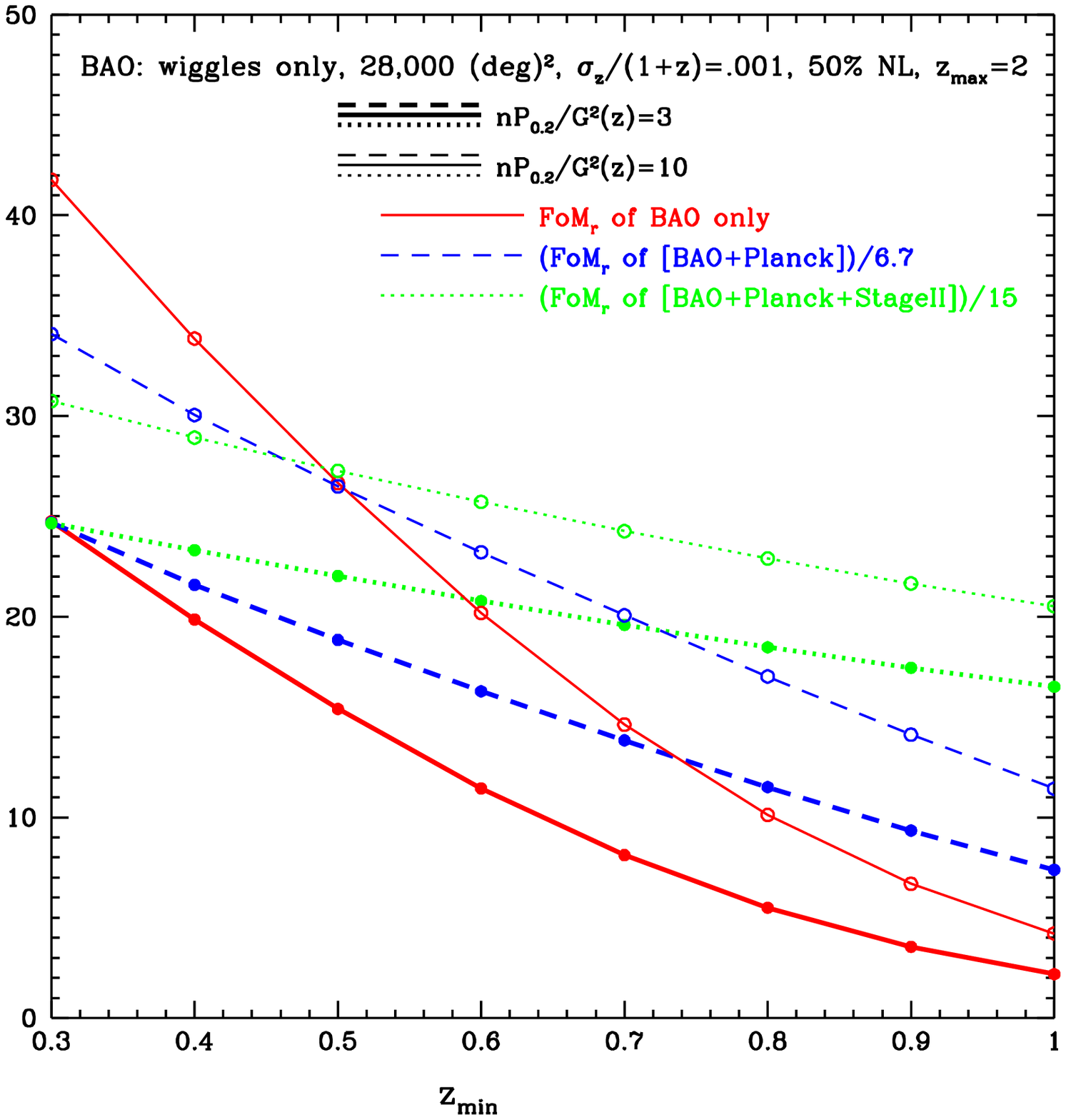,width=2.5in}
\caption{\label{fig:FoM_z}
\footnotesize%
The relative dark energy FoM, FoM$_r$, for BAO wiggles only from
a galaxy redshift survey
covering 28,000 (deg)$^2$ and $\sigma_z/(1+z)=.001$, as a function of 
$z_{max}$ for $z_{min}=0.5$ (upper panel), and as a function
of $z_{min}$ for $z_{max}=2$ (lower panel).
The DETF fiducial model is assumed.
}
\end{figure}

Fig.\ref{fig:FoM_area} shows the FoM for a galaxy redshift survey
covering $0.3<z<2.1$ with $\sigma_z/(1+z)=.001$, as a function of 
survey area. The BAO only constraints scale linearly with survey
area. This is as expected, since no priors are added, and the constraints
on dark energy parameters scale with $1/\sqrt{\mbox{area}}$.
The more external priors are added, the more slowly the FoM
increases with survey area. Thus the FoM of (BAO+Planck) increases
more slowly with area than that of BAO only, and the
FoM of (BAO+Planck+StageII) increases with area even more slowly.

\begin{figure} 
\psfig{file=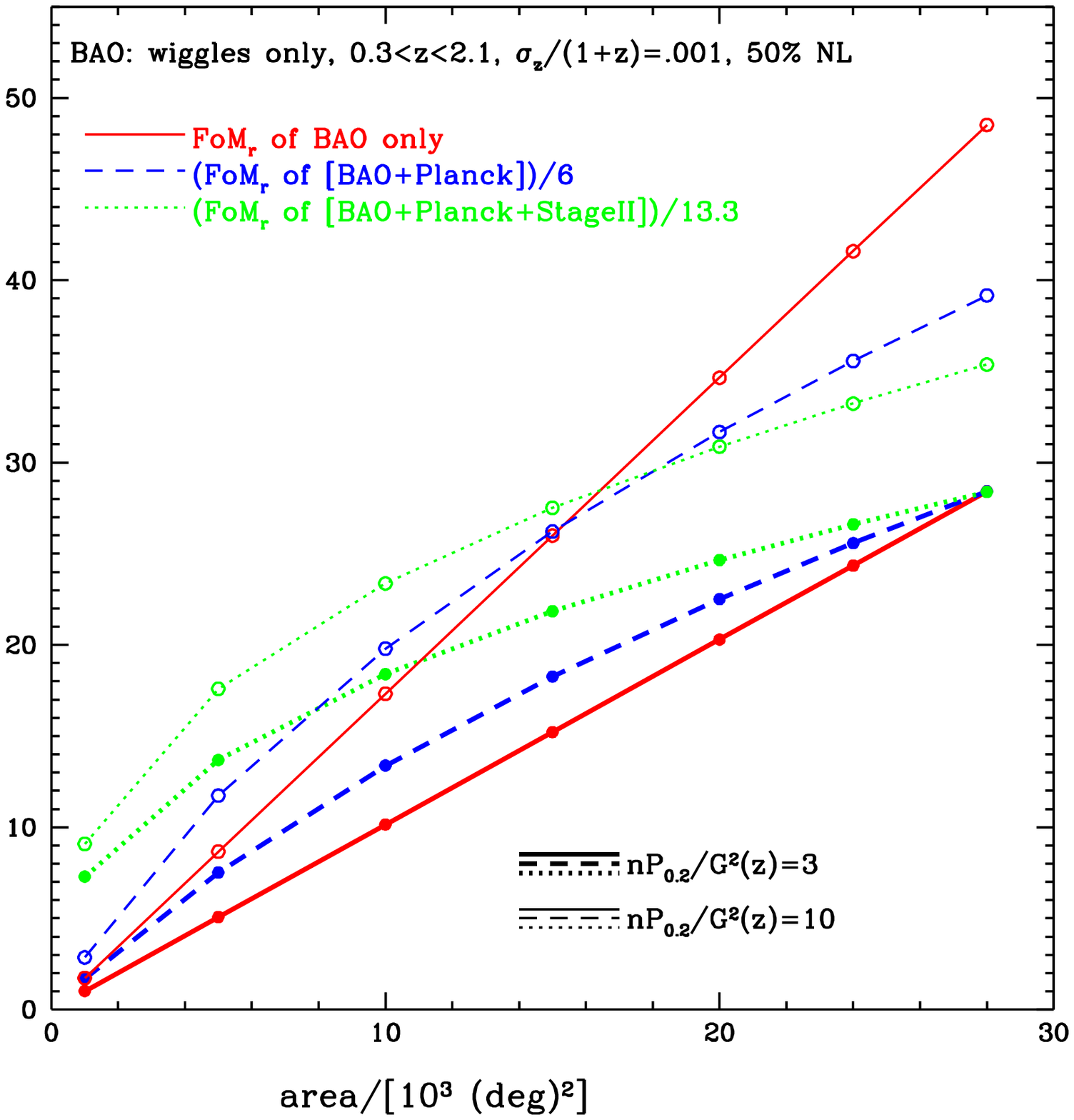,width=3.5in}
\caption{\label{fig:FoM_area}
\footnotesize%
The relative dark energy FoM, FoM$_r$, for BAO wiggles only from
a galaxy redshift survey
covering $0.3<z<2.1$ with $\sigma_z/(1+z)=.001$, as a function of 
survey area.
The DETF fiducial model is assumed.
}
\end{figure}

\subsection{Dependence on systematic uncertainties}

There are two types of systematic uncertainties for
BAO. One type of uncertainty is due to the calibration of
the sound horizon by CMB data. The other type of
uncertainty is due to the intrinsic limitations of the
BAO method, for example, the uncertainties
in the theory of nonlinear evolution and galaxy biasing
\cite{Angulo08}. 

WMAP five year observations gives $\Delta s/s \sim 1.3$\% 
\cite{Komatsu08}. We expect that Planck will give 
$\Delta s/s \sim 0.2$\% \cite{Planck}. Note that $s$ appears 
as an overall scale parameter in the observables $(s/D_A)$ and $(sH)$, 
and is assumed to be statistically independent from 
$(s/D_A)$ and $(sH)$. Thus the uncertainty in $s$
has {\it no} effect on the dark energy FoM
for the BAO wiggles only method. However, when other data are
combined with the BAO (wiggles only) data, the overall
dark energy FoM is decreased due to the uncertainty in
$s$. If the parameter set used is $(w_0, w_a, \Omega_X,
\Omega_k, \Omega_m h^2)$ for the BAO dark energy 
covariance matrix, then $\Delta s \neq 0$ only modifies
the diagonal matrix for $\Omega_m h^2$ by adding a
term
\be
\Delta C_{DE,55}^{BAO}= \left(2\Omega_m h^2 \right)^2 
\left(\Delta s/s\right)^2.
\ee
The Planck prior of $\Delta s/s \sim 0.2$\% has a very small
effect on the FoM. 

Our current modeling of intrinsic BAO systematic effects are at 
the $\sim$1-2\% level for realistic N-body simulations that include
galaxies (not just dark matter or dark matter haloes) \cite{Angulo08}. 
Correcting the bias in the estimated BAO scale due to systematic effects 
will likely lead to an increase in the uncertainty of the derived BAO 
scale \cite{Sanchez08}.
In order to make realistic assessment of dark energy constraints 
from BAO, one should allow for some level of remaining system uncertainty
in each redshift bin. 

We show the effect of systematic uncertainty in two different
ways. First, we show the effect of nonlinear effects explicitly,
using the parametrization shown in Eq.(\ref{eq:NL}), with
$p_{NL}$ indicating the level of remaining nonlinearity.
Secondly, we will show the effect of additive systematic
noise in each redshift slice due to the incomplete removal
or imperfect modeling of nonlinear effects or scale-dependent
bias.

Fig.\ref{fig:FoM_nP0} shows the relative dark energy FoM
for three levels of nonlinearity: 50\%, 75\%, and 100\%.
Clearly, dark energy constraints from BAO are extremely
sensitive to the level of nonlinearity assumed.
Fig.\ref{fig:FoM_NL} shows the FoM$_r$ as a function of
the level of nonlinearity for two representative galaxy number
distributions.

\begin{figure} 
\psfig{file=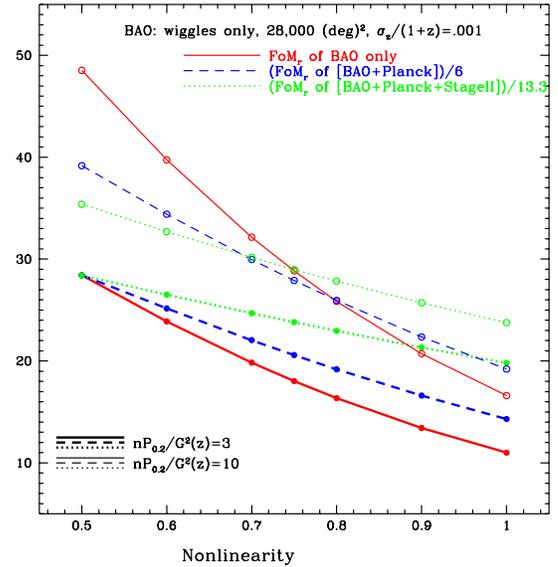,width=3.5in}
\caption{\label{fig:FoM_NL}
\footnotesize%
The relative dark energy FoM, FoM$_r$, for BAO wiggles only from
a galaxy redshift survey
covering 28,000 (deg)$^2$ and $0.3<z<2.1$, as a function of the level
of nonlinearity.
The DETF fiducial model is assumed.
}
\end{figure}

Fig.\ref{fig:FoM_dsys} shows the FoM$_r$, for a galaxy redshift survey
covering 28,000 (deg)$^2$ and $0.3<z<2.1$, as a function of the level
of additive systematic errors in each redshift bin.
The pessimistic case considered by the DETF, $\sigma_{sys}^i=0.01\times
\sqrt{5/\Delta z_i}$ gives 0.0224 for $\Delta z=0.1$, beyond the range
of Fig.\ref{fig:FoM_dsys}.
An additive systematic error of 0.5\% per $\Delta z=0.1$ redshift slice
would decrease the BAO FoM by 23\% for $nP_{0.2}/G^2(z)=3$,
and 34.5\% for $nP_{0.2}/G^2(z)=10$. Note however, the larger
galaxy number density should allow better modeling of the systematic
effects, thus the systematic errors should be lower for 
$nP_{0.2}/G^2(z)=10$. An additive systematic error of 0.35\% per 
$\Delta z=0.1$ redshift slice would decrease the BAO FoM by 21\% for 
$nP_{0.2}/G^2(z)=10$.

\begin{figure} 
\psfig{file=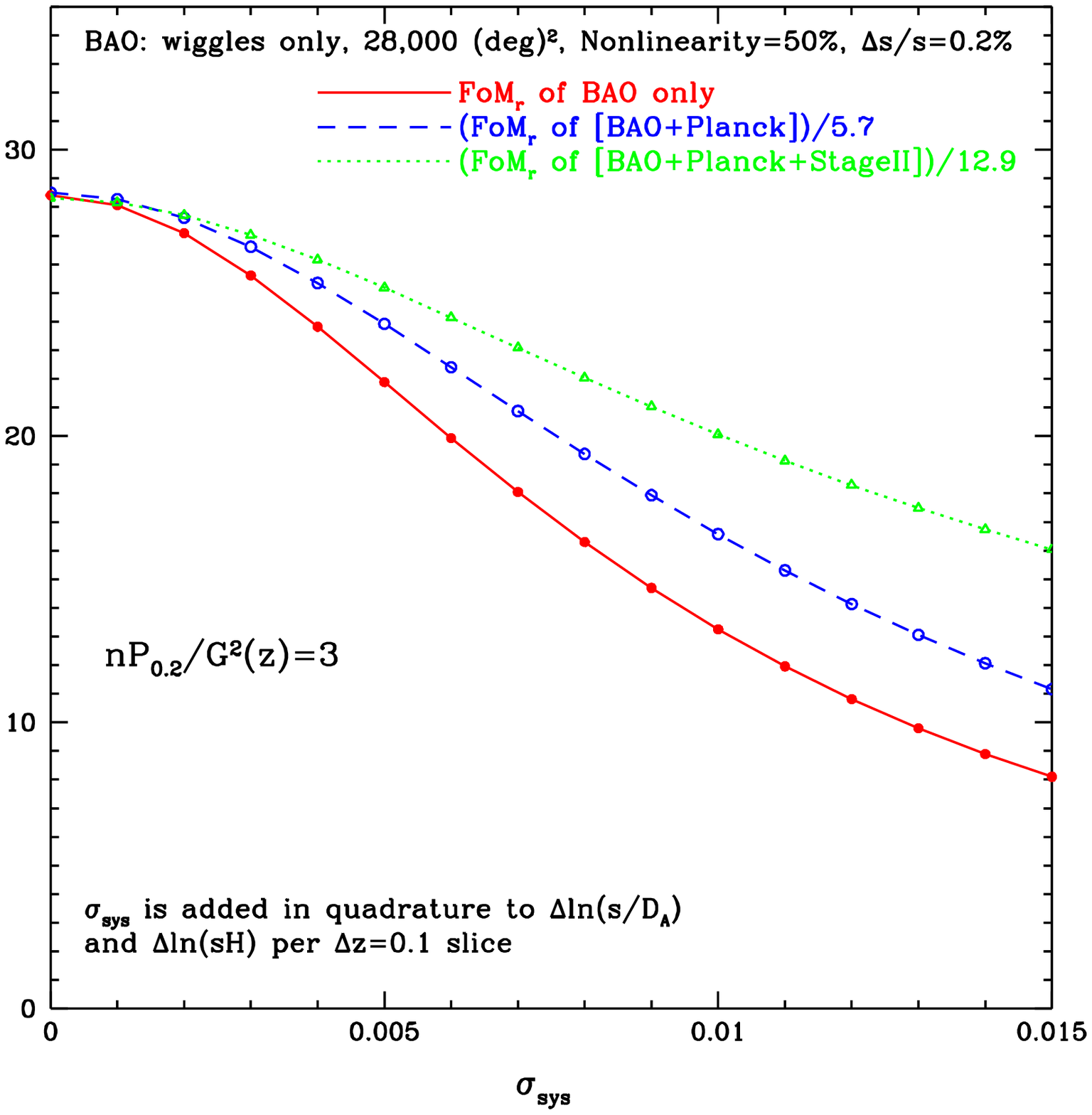,width=2.5in}
\psfig{file=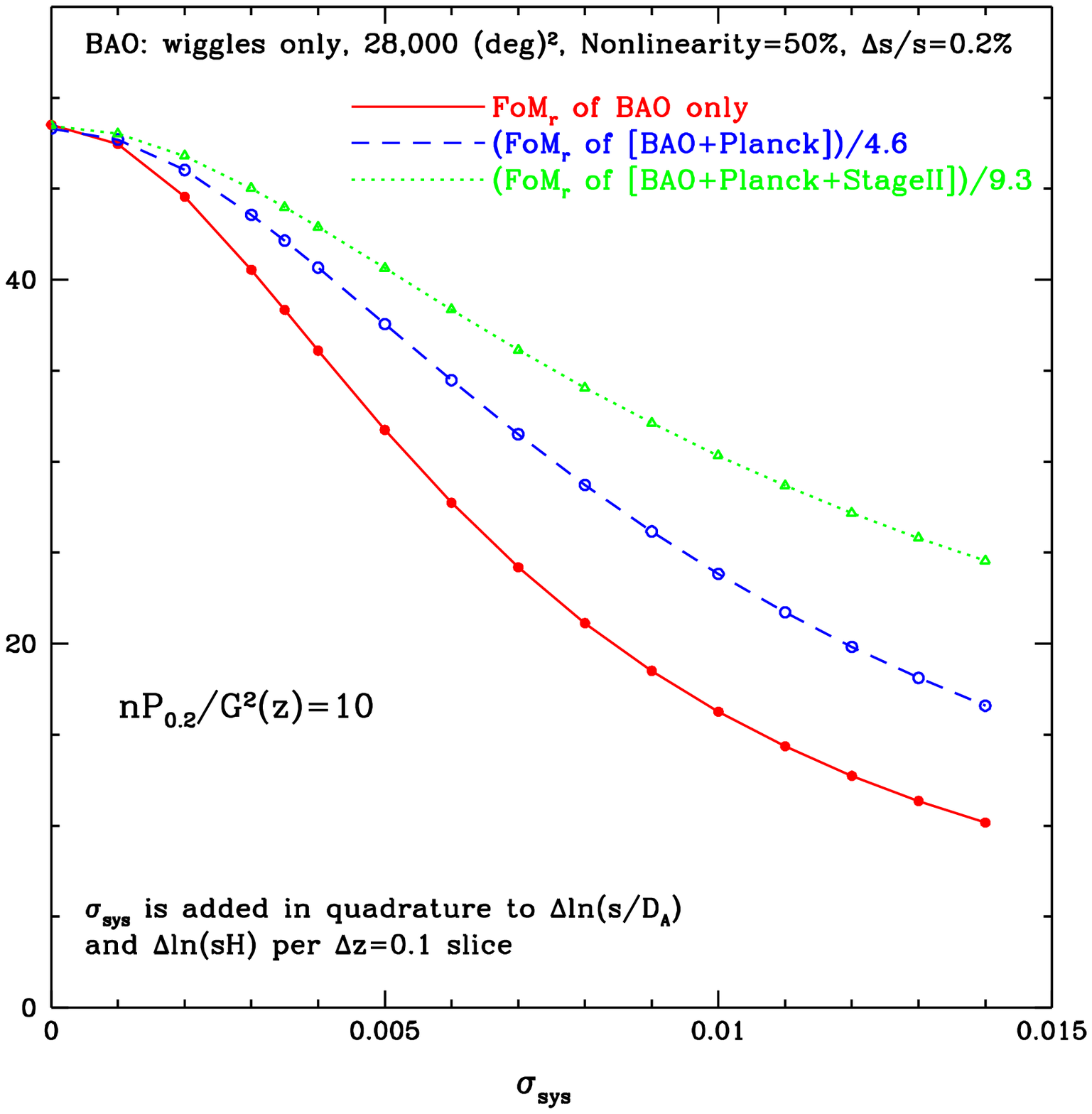,width=2.5in}
\caption{\label{fig:FoM_dsys}
\footnotesize%
The relative dark energy FoM, FoM$_r$, for BAO wiggles only from
a galaxy redshift survey
covering 28,000 (deg)$^2$ and $0.3<z<2.1$, as a function of the level
of systematic errors in each redshift bin.
The DETF fiducial model is assumed.
}
\end{figure}

\section{Summary}

We have studied the various assumptions that go into the BAO forecasts, 
using the fitting formulae for BAO (``wiggles only'') \cite{SE07}.
We have shown that assuming $nP_{0.2}/G^2(z)=\,$constant gives
a more realistic approximation of the observed galaxy number density
than $nP_{0.2}=\,$constant.
We find that assuming $nP_{0.2}/G^2(z)=10$ gives
very similar dark energy constraints to assuming $nP_{0.2}=3$, but
the latter corresponds to a galaxy number density larger by $\sim$
70\% at $z=2$ (see Fig.\ref{fig:nP}).

Assuming that $nP_{0.2}/G^2(z)=\,$constant, 
we have shown how the FoM for constraining dark energy depends 
on the assumed galaxy number density, redshift accuracy, redshift range, 
survey area, and the systematic errors due to uncertainties in the 
theory of nonlinear evolution and galaxy biasing.

We find that $\sigma_z/(1+z)=0.001$ is an optimal redshift accuracy for
a galaxy redshift survey, with the BAO FoM decreasing sharply with
increasing $\sigma_z/(1+z)$ for $\sigma_z/(1+z)>0.001$.
Further, the BAO FoM is very sensitive to the redshift range 
of the survey (see Fig.\ref{fig:FoM_z}), at both the low redshift
and high redshift ends. The optimal galaxy redshift survey should
measure the BAO using the {\it same} tracer over the entire redshift
range in which dark energy is important, i.e., $0 \la z \la 2$,
this would enable robust modeling of systematic effects, as well as
enabling the strongest dark energy constraints from BAO only.

The FoM of BAO is very sensitive to the level of nonlinearity assumed.
If the level of remaining nonlinearity is 80\%, instead
of the best case of 50\%, the FoM of BAO is decreased
by almost a factor of two (see Fig.\ref{fig:FoM_NL}).
The assumed sound horizon calibration error of $\Delta s/s=0.2$\%
(expected from Planck) has no effect on BAO only constraints,
and a negligible effect on the FoM of BAO combined with other data.
An additive systematic noise of up to $\sim 0.4-0.5$\% per $\Delta z=0.1$ 
redshift does not lead to significant decrease in the BAO FoM
(see Fig.\ref{fig:FoM_dsys}).

Finally, we note that future dark energy surveys are usually assessed 
by their performance when combined with Planck and Stage II priors \cite{detf}.
When this is done, it should be clearly stated.
Adding Planck priors to BAO typically boosts the FoM by about a factor of 
$\sim\,$5 compared to BAO only. Adding both Planck and DETF Stage II priors 
boosts the FoM by about a factor of $\sim\,$10 compared to BAO only 
(see Figs.\ref{fig:FoM_dlnz}-\ref{fig:FoM_dsys}). 
Fig.\ref{fig:FoM_dlnz}, Fig.\ref{fig:FoM_NL}, and Fig.\ref{fig:FoM_dsys} 
show that the FoM of combining BAO with Planck is less sensitive to
redshift errors and systematic errors. This is more so when
Stage II priors are added. The BAO FoM (BAO alone or combined with other data) 
also depend the choice of the fiducial cosmological model. 

A transparent comparison of the forecasts from different BAO projects
can only be achieved if all are required to choose (in the order of impact)
(1) the same BAO approximation method (for example, the wiggles 
only fitting formulae of Eq.[\ref{eq:Fisher_Wang}]);
(2) the same priors from Planck and current or ongoing projects (for example,
the DETF Stage II priors);
(3) the same fiducial model (for example, the
DETF model, or the five year WMAP bestfit model).

\bigskip

{\bf Acknowledgements}
I thank Daniel Eisenstein for clarifying SE07 and the BAO discussion in
Ref.\cite{FoMSWG}, and for useful comments.

\end{document}